# The SolarEV City Paradox: A Critical Review of the Fragmented Integration of Rooftop Photovoltaics and Electric Vehicles for Urban Decarbonization

Kobashi, T. [a],[*] R. C. Mouli [b], J. Liu [c], S. Chang [d], C. D. Harper [e], R. Zhou [c], G. R. Dewi [f], U. W. R. Siagian [f], J. Kang [g], P. P. Patankar [h], Z. H. Rather [h], K. Say [i], T. Zhang [a], K. Tanaka [j,k], P. Ciais [j], D. M. Kammen [l]

[a] Graduate School of Environmental Studies, Tohoku University, Sendai, Miyagi, Japan.
[b] Delft University of Technology, Delft, Netherlands.
[c] Harbin Institute of Technology, Shenzhen, Shenzhen, China.
[d] Purdue University, Purdue, USA.
[e] Carnegie Mellon University, Pittsburgh, Pennsylvania, U.S.A.
[f] Bandung Institute of Technology, Bandung, Indonesia.
[g] Seoul National University, Seoul, South Korea.
[h] Indian Institute of Technology Bombay, Mumbai, India.
[i] University of Melbourne, Melbourne, Australia
[j] Laboratoire des Sciences du Climat et de l'Environnement (LSCE), CEA, CNRS, UVSQ, Université Paris-Saclay, Gif-sur-Yvette, France
[k] Earth System Division, National Institute for Environmental Studies (NIES), Tsukuba, Japan
[l] Department of Civil and Systems Engineering (CASE) and Paul Pitze School of Advanced International Studies (SAIS), Johns Hopkins University, Baltimore MD/Washington DC, USA

[*]Corresponding author. Graduate School of Environmental Studies, Tohoku University, Sendai, Miyagi, Japan. *Email address*: takuro.kobashi.e5@tohoku.ac.jp (T.Kobashi).

**Abstract**

Urban decarbonization and an accelerating pace of innovation is pivotal to meeting global climate goals, yet progress toward fully integrated low-carbon energy systems remains inadequately slow. The *SolarEV City Concept*, which couples rooftop photovoltaic (PV) generation with electric vehicles (EVs) as mobile storage, offers a technically robust pathway to achieve deep $CO_2$ emission reductions, potentially meeting 60-95 % of municipal electricity demand when synergistically deployed. Despite rapid worldwide growth of both PV and EV markets, their integration through bidirectional Vehicle-to-Home (V2H) and Vehicle-to-Grid (V2G) applications has advanced slowly, revealing a persistent 'SolarEV paradox'. This review critically analyzes that paradox through a socio-technical framework encompassing four interdependent dimensions: technology, economics, policy, and society. Cross-national comparison shows that while technical feasibility is largely established, large-scale implementation is constrained by fragmented charging-protocol standards, immature and often non-profitable V2G business models, regulatory misalignments between energy and transport sectors, and social-equity barriers that limit participation to predominately high-income homeowners. Emerging national archetypes, from Japan's resilience-driven pioneer to Europe's regulation-first integrator, demonstrate that systemic integration trajectories are highly path-dependent at present. The analysis concludes that the transition toward SolarEV Cities requires a paradigm shift: moving beyond parallel promotion of PV and EV technologies toward coordinated policy frameworks, interoperable digital infrastructure, and inclusive market designs that equitably distribute both the economic and resilience benefits among citizens. Achieving this systemic integration will demand strategic orchestration among researchers, policymakers, industry, and communities to realize an adaptive, resilient, and socially just urban energy future.

*Key words*: Urban decarbonization, Rooftop photovoltaics, Electric vehicles, SolarEV city concept, Energy transitions, Socio-technical systems, Energy justice, Smart cities

Highlights
- Reveals the global 'SolarEV paradox' in PV–EV integration for decarbonization
- Applies a novel socio-technical framework across 4 systemic dimensions
- Compares 9 countries to identify divergent SolarEV City development archetypes
- Identifies critical barriers in standards, markets, policy misalignment, and equity gaps
- Synthesizes a paradigm shift toward integrated, equitable SolarEV City systems

1. **Introduction** (1000 words)

Anthropogenic climate change necessitates an urgent global transition away from fossil fuels [1,2]. Given that urban areas account for 67-72% of global energy-related $CO_2$ emissions, urban decarbonization represents a critical imperative for achieving international climate targets, such as those outlined in the Paris Agreement [2]. Beyond climate mitigation, decarbonizing coupled urban energy-transportation systems yields substantial co-benefits, including improved air quality, reduced noise pollution, and enhanced public health, contributing to more sustainable and livable cities and a means to invest in environmental equity and justice [3]. Despite this recognized urgency, effective, scalable, and equitable pathways for achieving deep urban decarbonization remain insufficiently developed and implemented globally [4].

In addressing the challenge of climate change, two emerging trends open the door to leveraging technological shift to accelerate urban carbon neutrality: the proliferation of distributed solar power generation, particularly rooftop photovoltaics (PV), and the electrification of transportation via electric vehicles (EVs) [5–7]. PV technology has emerged as the most cost-effective and rapidly scalable source of new electricity generation globally [6,8]. In 2024, global PV installations reached approximately 600 GW, contributing to a cumulative capacity exceeding 2.2 TW. However, PV's inherent intermittency limits its dispatchability (currently ~7% of global generation) and mandates the integration of energy storage solutions for reliable grid operation and deep decarbonization [6,9]. Rooftop PV systems, constituting a significant portion of global PV capacity (~40%), are particularly well-suited for urban environments [10]. They leverage existing building stock, minimize land-use conflicts associated with ground-mounted systems, and reduce transmission and distribution losses by co-locating generation with demand. As the Least Invasive Renewable Energy (LiRE) option, they offer potential for enhanced energy autonomy, while preserving natural ecosystems [11–15]. Studies indicate that maximizing technical potential of rooftop PV could significantly contribute to, or even meet, national electricity demands [7,16]. Recent debate, however, highlights uncertainties and emphasizes that national-scale estimates remain sensitive to methodological assumptions [17].

Concurrently, the global EV market is experiencing exponential growth, with global scales surpassing 17 million units in 2024 (>20% market share of passenger vehicles) [5]. Projections suggest EVs could dominate new vehicles sales by 2040 [18]. To fully realize the decarbonization potential of EVs, they must be powered by clean electricity sources and systematically integrated into the energy system as flexible assets [19]. Bidirectional charging technologies, collectively termed Vehicle-to-Everything (V2X), including Vehicle-to-Home (V2H), Vehicle-to-Building (V2B), and Vehicle-to-Grid (V2G), enable EVs to function as mobile energy storage units, capable of providing valuable grid services such as demand response, frequency regulation, and reserve capacity [20–23]. While the potential is significant, widespread V2X deployment currently remains limited, with notable exceptions primarily confined to specific applications like V2H for resilience in Japan [24].

The "SolarEV City Concept" arises from the theoretical synergy between these two rapidly evolving technologies [15,25]. It envisions future urban energy systems, where extensive rooftop PV generation is intelligently integrated with large fleets of EVs acting as distributed battery storage, thereby creating a more resilient, cost-effective, and deeply decarbonized energy infrastructure. Foundational research underscores the concept's substantial potential [15]. City-scale modeling indicates that optimizing the integration of rooftop PV (e.g., 70% rooftop area utilization) with EVs as battery could supply 60-95% of municipal electricity demands, a significant increase compared to the 30-40 % potential estimated for PV deployment alone [7]. Various case studies project considerable $CO_2$ emission reductions (42-95%) and energy cost savings (21-59%) across diverse urban contexts [7,15,26–29]. However, the performance and viability remain highly sensitive to local conditions, including solar irradiance patterns, existing grid carbon intensity, urban morphology, building stock characteristics, and prevailing policy environments [15,30,31]. Our Paris (Île-de-France) study showed that the suburbs operate as the metropolitan renewable backbone, effectively acting as the power plant for central Paris [30].

This review addresses a critical paradox characterizing the current state of the urban energy transition: despite the demonstrable potential of PV-EV integration and the rapid, independent market growth of both technologies, their systemic integration remains notably slow, fragmented, and hindered by numerous barriers. While existing literature often examines rooftop PV deployment, EV adoption trends, or V2G technological feasibility in isolation, a critical research gap exists concerning the synthesis of systemic interdependencies, cross-sectoral friction points, and the often-overlooked socio-political dimensions, including energy justice implications and behavioral challenges. This review aims to fill this gap by providing a comprehensive, critical analysis of the SolarEV City concept, moving beyond descriptive summaries to evaluate progress, persistent paradoxes, and potential pathways forward across diverse global settings. Specifically, this study seeks to answer the following research questions:

- What are the current technological, economic, and policy advancements enabling the systemic integration of rooftop PV with EVs in urban contexts, and where do critical misalignments persist across domains?
- Beyond greenhouse gas emission reductions, how does the SolarEV City model potentially reconfigure urban energy governance structures, and what are the implications for energy resilience, energy democracy, and social equity?
- What are the primary path-dependent and context-specific barriers (technical, market, regulatory, social) inhibiting the scaling of integrated SolarEV systems, and what distinct development archetypes are emerging globally?

To address these questions, this paper is structured as follows. Section 2 introduces the socio-technical framework employed as the analytical methodology. Section 3 examines the technological foundations and associated challenges, particularly concerning interoperability and infrastructure requirements. Section 4 examines the economic, policy, and social dimensions, analyzing barriers related to business models, regulatory frameworks, and social acceptance. Section 5 presents comparative analysis of global developments through case studies of nine selected countries, identifying divergent national pathways. Section 6 synthesizes the key challenges and future perspectives for scaling the concept. Finally, Section 7 provides concluding remarks and outlines an agenda for future research and policy intervention.

## 2. Analytical Framework: A Socio-Technical Approach

The "SolarEV City Concept" proposes the integration of rooftop PVs with EVs as batteries at the city scale (Fig. 1) [15]. This approach aims to mitigate the variability of PV output and decarbonize both the transport and building sectors simultaneously. Foundational techno-economic analyses demonstrate that this synergistic integration, assuming high (e.g., 70% of rooftop area) PV utilization, could supply 60-95% of municipal electricity demands, a significant increase compared to the 30-40% potential estimated for PV deployment alone (Fig. 2a,b) in Japan [7]. In addition, total annual PV generation surpasses total annual demand in many municipalities (Fig. 2c), creating opportunities for surplus renewable energy to be utilized for other applications, such as green hydrogen production or export to other regions [7].

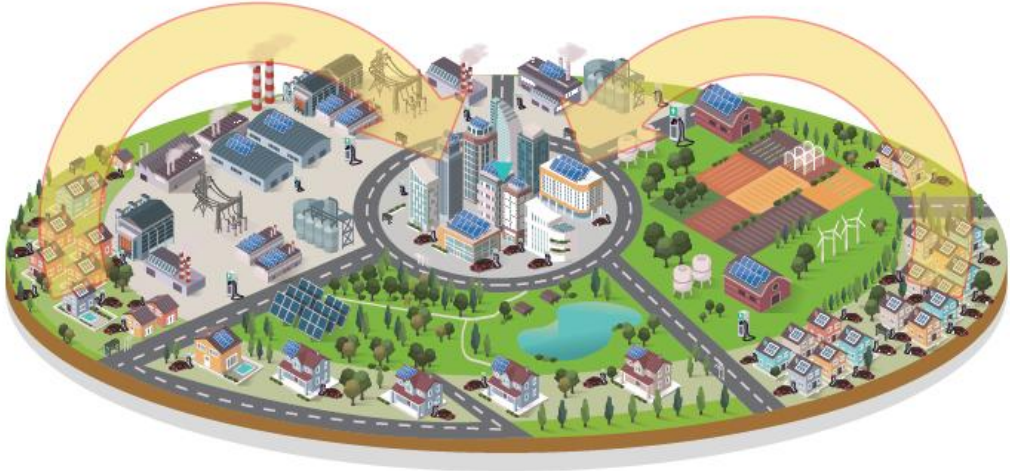

Figure 1. A conceptual image of the *SolarEV City*. Rooftop PV + EV systems supplying the city's energy demand. Realization requires alignment across four interdependent pillars: Technology, Economics, Policy, and Society. The two arrows indicate electricity is supplied to the demand center from surrounding suburbs.

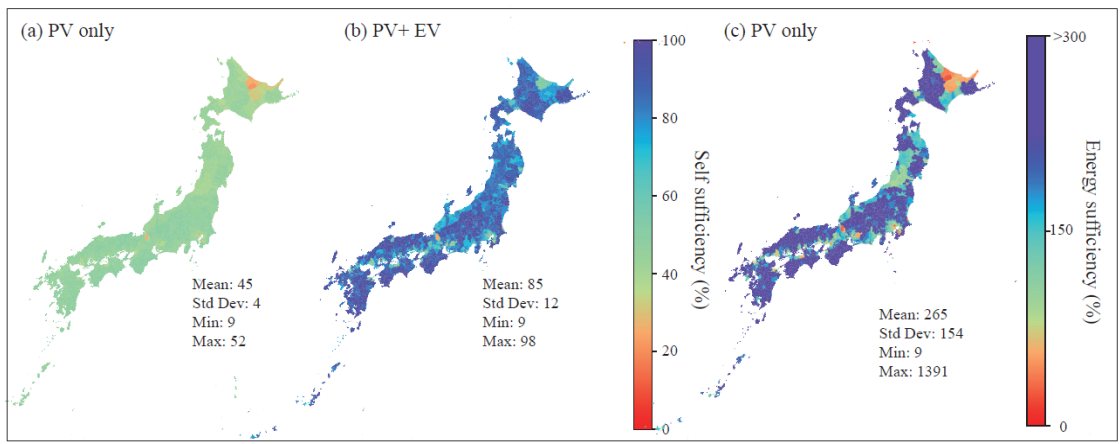

Figure 2. Decarbonization potentials of rooftop PV integrated with EV systems for all the municipalities of Japan [7]. (a) Self-sufficiency (total locally consumed PV electricity/total demand) of rooftop "PV only" system, (b) Self-sufficiency of rooftop PV integrated with EVs, (c) Energy sufficiency (total PV generation/demand) of rooftop "PV only" system [7]. Mean, Standard Deviation (Std Dev), Minimum (Min), and Maximum (Max) were calculated from all the 1741 municipalities of Japan [7]. The systems assume the utilization of 70% rooftop area for PV and all the passenger vehicles as EV (as flexibility: i.e., 20 kWh/vehicle) in the municipalities [7].

The transition towards a SolarEV City (Fig. 1) represents a complex socio-technical transformation, involving a fundamental reconfiguration of urban energy systems rather than mere technological substitution [25,32–35]. Critically analyzing the progress, impediments, and inherent paradoxes, specifically the rapid individual deployment versus slow systemic integration, necessitates a holistic methodology beyond isolated assessments. Therefore, this review employs a socio-technical system perspective, conceptualizing the *SolarEV City* as a system requiring alignment across four interdependent pillars: 'Technology', 'Economics', 'Policy', and 'Society' (Fig. 3). This methodological framework suggests that successful integration hinges on coherence at the interfaces between these pillars, while barriers arise from friction, misalignment, or gaps between them. It provides the structured lens applied throughout this review to diagnose systemic issues and analyze divergent global pathways.

Applying this framework, our analysis first dissects the Technology pillar, examining core enablers like rooftop PV (prioritizing its urban advantages such as utilizing built surfaces [36,37], enhancing grid resilience [38], minimizing losses [39], fostering energy democracy [40,41]), and EVs as mobile storage through V2X technologies and their aggregation into Virtual Power Plants (VPPs). Critically, we assess integration challenges such as grid impacts, infrastructure constraints, and the fundamental barrier of interoperability arising from fragmented charging standards ("protocol" challenges). Subsequently, the "Economics" pillar analysis evaluates value creation mechanisms (both micro-level self-consumption and macro-level market participation via V2G) and assesses the viability and scalability of emerging business models like VPPs and Peer-to-Peer (P2P) trading, alongside significant economic barriers such as high upfront costs.

Finally, the methodology utilizes the "Policy" pillar to investigate governance structures, regulatory gaps (particularly between energy and transport sectors), market access rules for distributed resources, and the integration of renewable/EV mandates into urban planning. The "Society" pillar analysis focuses on critical social dimensions, applying an energy justice lens to evaluate equity implications (access for renters, Low-to-Moderate Income (LMI) households) [42] and examining social acceptance issues, including behavioral barriers related to battery degradation concerns, data privacy, and trust [43]. This comprehensive socio-technical approach guides the subsequent examination of technological foundations, economic, and policy dimensions, and future challenges. Crucially, this framework is systematically applied in Section 5 to compare the distinct national pathways and development archetypes emerging across the selected case countries.

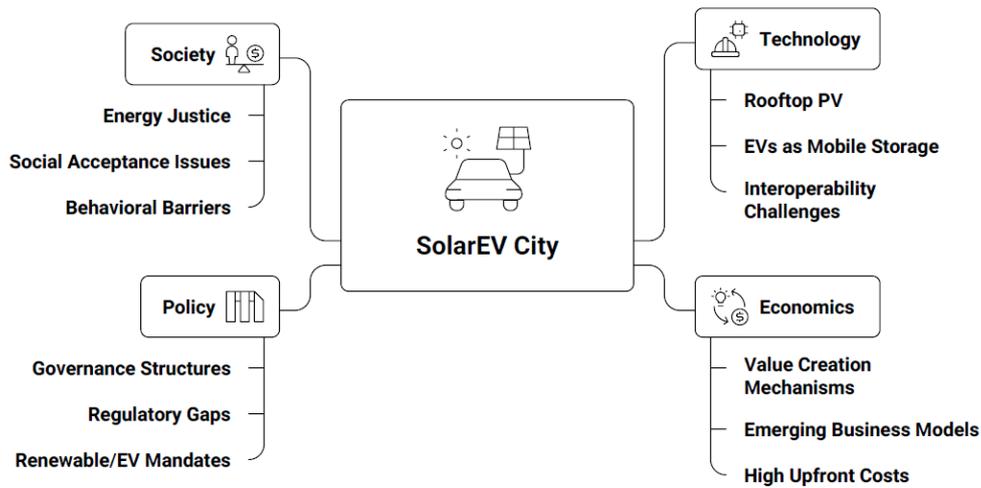

Figure 3. Analytic framework of realizing the *SolarEV City*.

## 3. Technological Foundations and Challenges

Realizing the *SolarEV City* vision hinges upon the seamless integration of advanced PV systems and EV infrastructure within intricate urban energy ecosystems [15]. While considerable technological progress underpins this concept, formidable challenges pertaining to urban constraints, grid integration limitations, and particularly standardization must be rigorously addressed to unlock its full decarbonization potential.

3.1 Urban Solar Power Generation

The widespread deployment of PV systems within urban environments forms a foundational element of the SolarEV City concept, driven by significant technological innovation and increasingly supportive policy landscapes. High-efficiency monocrystalline silicon cells now surpass 25% conversion rates, while emerging technologies like perovskite solar cells hold promise for further performance gains (e.g., weight, flexibility, and efficiency) and cost reductions [44]. Crucially for space-constrained cities, Building-integrated Photovoltaics (BIPV), including transparent PV glass and façade-integrated modules, enable energy generation to be embedded directly into the building envelope, maximizing potential generation particularly on high-rise structures [45]. Concurrently, supportive policy frameworks, including renewable energy targets, financial incentives, and the integration of GIS-based solar potential assessments into urban planning, have spurred deployment [46]. Furthermore, innovative deployment models like community solar offer pathways to broaden access in dense urban areas [42]. These trends facilitate the synergy sought by the SolarEV City concept, where distributed PV generation can directly support EV charging infrastructure and contribute to urban carbon neutrality goals [25].

Despite these positive developments, urban PV deployment confronts substantial challenges inherent to the built environment. High population and building density severely limit available unshaded space for optimal PV installation, with shading from adjacent structures significantly reducing energy yields and complicating site selection (Horváth et al., 2016). Economic barriers remain significant, particularly the high upfront capital costs associated with installations, especially for retrofitting existing buildings where structural modifications or complex installation may be required, often compounded by elevated urban property and labor costs[47]. Technical and regulatory hurdles further impede progress; the frequent lack of granular, publicly accessible rooftop data and the computational expense of detailed solar potential assessments hinder precise planning, while permitting processes are often prolonged and may not adequately accommodate novel BIPV solutions [45,46].

A fundamental technical challenge intrinsic to solar power is its intermittent and variable weather dependent nature, driven by diurnal cycles and weather variability, which complicates its integration into urban energy grids that require stable supply for buildings and EV charging. Urban microclimates, characterized by complex shading and reflection patterns from buildings, exacerbate this variability and introduce significant forecasting uncertainties[48]. While geographical aggregation across urban areas can partially mitigate localized fluctuations, it necessitates coordinated infrastructure [49]. Consequently, accurate, high-resolution forecasting models, increasingly leveraging machine learning and real-time data, are indispensable for grid management [50]. Furthermore, high penetration levels of distributed PV can induce power quality issues (e.g., voltage fluctuations) within local distribution networks, requiring advanced inverter functionalities (e.g., Volt/VAR control) and integrated energy storage solutions like Battery Energy Storage Systems (BESS) to maintain grid stability [51,52]. Within the SolarEV City framework, BESS, potentially integrated with EV charging infrastructure, alongside demand response mechanisms (e.g., TOU tariffs) and overarching smart grid technologies (enabling real-time monitoring, predictive analytics, and dynamic load balancing) are essential for managing variability, ensuring reliability, and enabling the scalability of urban solar integration [25,53,54].

3.2 Electric Vehicles as Distributed Energy Resources

The capacity to integrate EVs not merely as passive loads but as active Distributed Energy Resources (DERs) is fundamental to the SolarEV City concept, offering crucial flexibility to manage the intermittency of rooftop PV generation and enhance overall grid stability. This integration leverages the inherent synergy between PV systems and EVs: EVs when connected to a suitable charger can store surplus solar energy and provide dispatchable power, while PV can offer a direct renewable source for vehicle charging. The key mechanisms to enable this integration is bidirectional charging functionalities, namely Vehicle-to-Grid (V2G), Vehicle-to-Home (V2H), and Vehicle-to-Building (V2B), which allow controlled power export from the EV battery, alongside intelligently managed unidirectional charging, and is often termed smart charging or V1G. These capabilities transform EVs into mobile energy storage assets capable of participating dynamically within the energy system. A recent milestone is Nissan's announcement in 2024 of an affordable 'on-board' bidirectional charging system enabling Vehicle-to-Home (V2H) and Vehicle-to-Grid (V2G) functionality from 2026, marking a critical step toward the mass-market adoption of integrated 'PV + EV' systems [55].

Research extensively documents the potential of EVs as DERs. Utilizing EV batteries to store excess solar generation may effectively mitigates PV intermittency, with studies demonstrating significant reductions on grid-sourced electricity, with a potential reduction of 64%-100% under certain conditions [56–58]. Furthermore, aggregated EV fleets when managed as VPPs, can provide substantial grid support during peak demand periods across residential and commercial sectors [29,59,60]. Algorithmic management of commercial EV fleets has shown potential to reduce industrial peak loads by as much as 50% and associated energy costs by 27% [61]. Increasingly, V2H and V2B applications are also being evaluated for their capacity to provide building-level services, including backup power during outages, demand-response load shaving, and optimization of self-consumption from onsite renewables [62–67].

Smart charging (V1G) strategies represent another critical mechanism, aligning EV charging schedules with grid conditions (e.g., low prices, high renewable availability) or local PV generation profiles. Coordination algorithms developed at both small-scale (e.g., parking lots with solar carports) and city-scale aim to maximize the utilization of solar energy for EV charging [68–73]. Modeling suggests that coordinated charging could supply a substantial fraction (13-92%) of EV energy demand directly from solar, depending on factors like coordination level and seasonality [71]. Moreover, scheduling charging for off-peak hours and utilizing bidirectional discharging during peak periods can significantly reduce peak electricity demand (with potential reductions up to 44%), thereby alleviating stress on distribution networks [74–77]. Consequently, the large-scale, intelligent integration of EVs as DERs, particularly when coupled with distributed PV, presents significant opportunities for optimizing grid planning, enhancing reliability, and accelerating the transition towards decarbonized urban energy and mobility systems [78–82].

3.2.1. The Protocol Challenge: Interoperability and Standardization

A critical, yet frequently underestimated, technical barrier impeding the widespread adoption and scaling of V2G functionalities is the persistent fragmentation of global charging communication standards. The ability for an EV to communicate bidirectionally with charging infrastructure (Electric Vehicle Supply Equipment, EVSE) is governed by specific protocols, and the absence of a single, universally adopted, and fully implemented standard creates significant interoperability challenges. This lack of standardization hinders the development of a seamless, cost-effective, and scalable global market for V2G services, representing a major friction point within the technology pillar. Japan's relative success in establishing a commercial V2H market can be directly attributed to its early standardization around the CHAdeMO protocol [83], which inherently supported bidirectional power flow, driven by national resilience objectives in the aftermath of Fukushima.

Conversely, Europe and North America largely converged on the Combined Charging System (CCS) standard, which initially prioritized high-power unidirectional charging to address consumer range anxiety, deferring the incorporation of the bidirectional capacity to later development stages. The primary standard enabling bidirectional communication and advanced V2G functionalities for CCS, ISO 15118 (specifically the ISO 15118-20-edition supporting bidirectional power transfer), has only recently reached maturity, with widespread commercial implementation in vehicles and chargers still emerging [84]. This significant "protocol lag" has resulted in a path-dependent problem: millions of existing CCS-compatible EVs lack the necessary hardware and software for V2G participation, creating a substantial legacy fleet unable to provide grid services and delaying the aggregation potential required for VPPs to achieve critical mass. Further complicating the landscape is the joint China–Japan development of the ChaoJi high-power bidirectional charging system as an evolution of the GB/T and CHAdeMO standards [85], which adds another layer of complexity for global automakers and infrastructure providers pursuing interoperability. It is noted that Tesla's Supercharger and vehicle architecture currently do not support bidirectional power flow, underscoring that even leading EV automakers have yet to implement full V2G capability within their charging ecosystems [86].

3.3. Smart Grids and Digital Infrastructure

The orchestration of numerous distributed PV systems and EV assets within a SolarEV City necessitates an advanced digital infrastructure layer, commonly referred to as a smart grid [87]. Smart grids leverage bidirectional communication, advanced metering, sensing, data analytics, and automated control systems to dynamically manage energy flows, ensure grid reliability, optimize resource utilization, and enable the seamless integration of diverse DERs [53,54]. Central to the functionality of this digital infrastructure are enabling technologies such as Artificial Intelligence (AI), which plays a pivotal role in managing the inherent complexity introduced by variable renewable generation and dynamic EV loads. AI, particularly machine learning (ML) and deep learning algorithms, facilitates high-accuracy electricity demand forecasting by analyzing complex datasets encompassing historic consumption, weather patterns, and urban activity trends [88]. Furthermore, AI-driven optimization techniques, including reinforcement learning, enable real-time, autonomous decision-making for maximizing grid efficiency, dynamically adjusting the operation of BESS and EV charging/discharging schedules to align with PV generation or grid needs, and facilitating sophisticated demand response programs [25,89,90].

Beyond AI, blockchain technology presents a potential framework for enhancing the efficiency, security, and transparency of energy transactions within a decentralized SolarEV City ecosystem. Blockchain offers a distributed ledger technology that can facilitate secure and transparent peer-to-peer (P2P) energy trading, enabling prosumers (e.g., households with surplus rooftop PV) to transact directly with consumers (e.g., EV owners) via automated smart contracts [91]. This model has the potential to foster local energy markets, empower consumers, and reduce reliance on centralized intermediaries. Blockchain applications also extend to tracking the origin of renewable energy used for EV charging, thereby enhancing the verifiable sustainability of electric mobility [92]. While pilot projects have demonstrated the technical feasibility of blockchain-enabled P2P trading, ongoing challenges related to scalability, transaction speed, standardization, and regulatory acceptance need to be addressed for widespread implementation [93].

The increasing digitalization and interconnectedness fundamental to smart grids, AI-driven energy management, and V2G communication inevitably create a significantly expanded cybersecurity attack surface. Securely managing the flow of sensitive operational and transactional data between millions of DERs, aggregators, grid operators, and consumers is paramount. Potential vulnerabilities could be exploited to disrupt charging operations, manipulate energy markets, compromise user privacy, or even destabilize the grid through coordinated attacks on aggregated VPPs. Therefore, embedding robust, multi-layered cybersecurity measures, including end-to-end encryption, strong authentication protocols (such as those specified in ISO15118), intrusion detection systems, secure software development practices, and resilient network architectures, is not an optional add-on but a fundamental prerequisite for the safe, reliable, and trustworthy operation of SolarEV City infrastructure [94].

4. Economic, Policy, and Social Dimensions

While technological advancements provide the necessary foundation, the successful transition towards scalable SolarEV Cities is fundamentally contingent upon navigating and resolving critical challenges within the economic, policy, and social domains. Analysis reveals that the most significant impediments frequently lie not in technical feasibility but in the lack of viable economic structures, persistent regulatory misalignments, and deep-seated social barriers, including important equity considerations.

4.1 Economic Feasibility and Business Models

At the microeconomic level, the primary incentive for individual adoption remains potential cost savings derived from maximizing the self-consumption of rooftop solar PV generation, particularly in regions with time-differentiated electricity tariffs. However, the substantial upfront capital investment required, including PV systems, EVs, and the significant cost premium associated with bidirectional EVSE, constitutes a prohibitive barrier for a large segment of the population (although significant future cost reduction is expected [25]). Consequently, achieving mass-market economic viability necessitates the development and "stacking" of multiple value streams beyond simple bill savings, primarily through participation in broader energy markets.

Aggregating the distributed capacity of numerous EVs through VPPs or facilitating P2P energy trading represents the most promising pathways for unlocking additional economic value. VPPs leverage sophisticated software platforms to orchestrate potentially thousands of DERs, including EV fleets, bidding their collective capacity into wholesale energy, capacity, or ancillary service markets, thereby generating revenue streams shared among participants (typically, the aggregator and EV owners). P2P trading platforms, often utilizing blockchain, aim to create localized energy marketplaces, potentially offering more favorable transaction prices by bypassing traditional utility structures [92,93,95]. However, both VPP and P2P models currently face substantial hurdles related to, gaining equitable market access for small-scale distributed resources, establishing fair and effective compensation mechanisms, defining the roles and responsibilities of new market actors like aggregators, and overcoming regulatory inertia, which highlights the critical friction between the economic and policy pillars.

4.2 Policy Frameworks and Governance

Policy and regulatory frameworks represent the most critical determinants governing the pace and trajectory of the SolarEV City transition. Historically, however, policy support for rooftop PV deployment and EV adoption has evolved largely along separate, parallel tracks, creating a significant impediment to their synergistic integration. Global rooftop PV policies have progressed through distinct phases: an initial policy-driven phase (pre-2017) dominated by mechanisms like feed-in tariffs (FITs) and subsidies (e.g., China, Germany, France) or tax credits and net metering (e.g., US) [96–100]; followed by a transitional phase (2018-2021) characterized by subsidy reductions driven by falling PV costs [101]; leading to the current market-driven phase emphasizing market participation, dynamic pricing, and blended policy portfolios [102–105]. Similarly, global EV policy has shifted from a primary reliance on direct purchase subsidies towards a more diversified toolkit including non-fiscal incentives (e.g., license exemptions in China), charging infrastructure support, manufacturer mandates (e.g., ZEV mandates in UK/US states), industrial policy alignment (e.g., Germany), and carbon-linked incentives (e.g., France) [106–112].

This historical development along separate tracks has resulted in a critical "regulatory gap" or "policy void" concerning the integrated operation of PV and EVs, particularly regarding V2G functionalities. V2G implementation resides uneasily at the intersection of transportation policy (governing vehicle standards and markets) and energy policy (governing grid interconnection and electricity markets), domains often managed by different agencies with distinct mandates and regulatory cultures. This gap manifests in several acute challenges. First, existing grid interconnection standards were predominantly designed for stationary generators, lacking clear, streamlined, and cost-effective procedures for connecting millions of mobile, bidirectional storage devices (EVs) [113]. Second, utilities often lack established protocols or business models for V2G, raising valid concerns about potential impacts on grid safety, stability, and power quality which can lead to cautious and complex approval processes, or a lack of clear rules for fairly compensating EV owners or aggregators for the diverse suite of grid services V2G can provide (ranging from fast frequency response to capacity provision). Third, discharging electricity back to the grid can potentially conflict with existing retail electricity contracts or create ambiguous taxation implications, thereby eroding the economic incentive for participation.

As EV adoption matures, policymakers globally are beginning to address this regulatory gap, albeit through diverse and often fragmented approaches, particularly concerning V2G integration. China exemplifies a top-down, pilot-driven model, enforcing national standards while initiating large-scale trials augmented by specific local incentives [114–116]. The US combines federal initiatives (e.g., C-V2X mandates) with pioneering yet inconsistent, state-level regulations (e.g., California's bidirectional mandate) within a highly fragmented utility landscape [117,118]. Germany employs a hybrid approach, mandating grid-controllable charging capabilities while providing premium subsidies for bidirectional hardware [119,120]. The UK focuses on enabling market participation through tariff structures (TOU rates for AC V2G) [121], while France leverages enterprise alliances within a more centralized energy system structure [122,123]. This ongoing regulatory uncertainty and lack of harmonized frameworks across jurisdictions remain as primary factors contributing to the slow commercial rollout and scaling of V2G services globally and represent a key bottleneck despite increasing technological readiness.

4.3 Social Dimensions: Behavior, Acceptance, and Justice

Overlooking the social dimension risks undermining the long-term viability and social legitimacy of the SolarEV City transition. Consumer adoption of V2G, even with financial incentives, faces significant behavioral barriers. Concerns regarding the potential acceleration of battery degradation and voiding of manufacturer warranties remain paramount among potential users [43,124]. While research suggests intelligent V2G management could potentially mitigate net degradation by optimizing battery state-of-charge and reducing calendar aging, clear communication and robust warranty coverage explicitly including V2G participation are crucial to alleviate these fears [125–128]. Additionally, perceived loss of control over personal mobility (range anxiety if the battery is discharged for grid services) and data privacy concerns associated with sharing charging data require user-centric control interfaces and trustworthy data governance frameworks [43].

Perhaps the most profound challenge lies in ensuring the transition aligns with the principles of energy justice [129]. The predominant SolarEV City model, often centered on single-family homeowners with private driveways and sufficient capital, risks systematically excluding renters, residents of multi-unit dwellings, and low-to-moderate households, thereby potentially exacerbating existing energy inequities [42]. This could lead to a socially stratified, "Two-tiered" energy system where affluent "prosumers" capture the benefits (lower bills, V2G revenues, resilience) while vulnerable populations remain reliant on a potentially more strained public grid and may bear a disproportionate share of system upgrade costs. Addressing this critical "equity gap" necessitates proactive policy design focused on inclusive models, such as expanding access to community solar programs, prioritizing the deployment of publicly accessible V2G-enabled charging infrastructure in underserved communities, developing innovative financing mechanisms for low-to-moderate households and multi-unit dwellings residents, and ensuring procedural, distributive, and recognition justice principles [42,130,131]. Achieving a truly sustainable SolarEV City requires embedding equity considerations at the core of planning and implementation.

## 5. Global Developments: A Comparative Analysis of Divergent Pathways

The development trajectory of the SolarEV City concept is markedly heterogeneous across the globe (Table 1), reflecting diverse national energy system characteristics, policy priorities, market structures, technological legacies, and socio-economic contexts. Rather than converging towards a single implementation model, distinct national pathways are emerging. This section presents a comparative analysis across nine key countries, applying the socio-technical framework to assess the status of rooftop PV deployment, EV market penetration, and critically, the progress and challenges associated with V2H/V2G integration. This cross-case synthesis reveals distinct development archetypes and underscores the persistent global paradox of fragmented integration despite rapid individual technology adoption.

Table 1. SolarEV City developments in various countries. EV includes Battery Electric Vehicles (BEV) and Plug-in Hybrid Electric Vehicles (PHEV).

| | Population (million) in 2024 [132] | Total PV installed capacity per | PV electricity penetration in total | EV share in the market | V2H protocols | V2H/V2G market development |
|---|---|---|---|---|---|---|

| | capita (kW) in 2024 [133] | generation in 2024 [134] | in 2024 [135] | | |
|---|---|---|---|---|---|
| Australia | 27 | 3.7 | 18.9% | 9.8% | CCS2 Piloting, CHAdeMO available | Pilot |
| China | 1,409 | 2.4 | 8.7% | 48.1% | GB/T Piloting, ChaoJi in development | Pilot |
| France | 69 | 2.2 | 4.2% | 27.3% | CCS2 (ISO 15118) Commercial | Commercial |
| India | 1,451 | 0.3 | 0.7% | 2.6% | CCS2/CHAdeMO Piloting | Nascent |
| Indonesia | 284 | 0.4 | 0.4% | 6.6% | CCS2/CHAdeMO Piloting | Nascent |
| Japan | 124 | 2.4 | 11% | 3.0% | CHAdeMO Commercial | Commercial (V2H) |
| Netherlands | 18 | 3.5 | 17.1% | 48.9% | CCS2 (ISO 15118) Commercial | Commercial |
| South Korea | 52 | 3.0 | 6.1% | 10.7% | CCS2/CHAdeMO Piloting | Pilot |
| U.S. | 340 | 4.1 | 6.5% | 10.6 % | CCS2 Piloting, CHAdeMO available | Pilot/Early commercial |

5.1 Australia

Australia presents a unique energy transition context characterized by world-leading per capita adoption of rooftop solar, driven by excellent solar resources and relatively high retail electricity prices, alongside a structural shift away from aging coal-fired generation [136]. This massive penetration of distributed PV (>40% of houses, 37.8 GW capacity) significantly impacts wholesale electricity markets, frequently causing very low or negative prices during midday solar peaks, which consequently diminishes the economic value of simple solar exports and creates strong incentives for enhancing self-consumption through energy storage [137,138]. While household battery adoption is increasing (~7.7% attachment rate), the EV market, although growing (10% new sales share in 2024), lags significantly behind PV deployment (Table 1). This disparity is partly attributable to policy uncertainty and relatively underdeveloped public charging infrastructure compared to global leaders [139–142].

This specific constellation of factors positions Australia within the 'Market-Led, Fragmented Innovator' archetype. There is immense bottom-up market pressure and consumer demand for solutions (like V2H/V2G) that maximize the value of existing rooftop PV assets, yet actual progress towards systemic integration remains paradoxically slow and largely confined to pilot projects. Bidirectional charging technology is nascent, transitioning from trials to early commercialization only recently and primarily hindered by the historical dominance of the CCS2 charging standard (which lacked bidirectional capability until recent protocol updates) and a complex, slow-moving regulatory process for establishing market participation rules, grid connection standards, and appropriate tariff structures for V2X services [143–145]. Although enabling legislation was passed in late 2024 and policies are gradually being loosened to allow household participation in energy markets, the lack of overarching policy coherence and coordinated market design means that realizing the full potential of PV-EV integration remains fragmented and uncertain, failing to fully leverage the nation's vast distributed energy resource base [146–149]. Consumer expectations however are increasingly shifting towards integrated behind-the-meter systems, driving the foundations of the SolarEV City concept.

5.2 China

China's energy landscape is defined by its unparalleled scale and strong centralized policy direction aimed at achieving ambitious "dual carbon" targets (peak 2030, neutrality 2060) (Fig. 4). It dominates global supply chains and deployment figures for PV, with cumulative installed capacity surpassing 1,080 GW by May 2025, accounting for nearly half of the global market [150]. This rapid expansion, which achieved the 2030 national wind and solar target six years early, has transitioned from an initial reliance on feed-in-tariffs [151] to more market-driven mechanisms [152]. Since 2021, distributed PV has consistently accounted for over half of new installations, with recent growth dominated by the industrial and commercial sectors (>70% of new distributed capacity in 2024) [153], which benefit from high electricity prices and clear investment returns, outpacing earlier rural-focused initiatives.

Concurrently, China is the undisputed global leader in the EV market, maintaining the top sales position for ten consecutive years [154]. In 2024, sales of new energy vehicles reached 18.9 million units, capturing 70.5% of the global market share [155]. Domestic market penetration is accelerating rapidly, projected to exceed the 50% target by 2025 or 2026, nearly a decade ahead of the original 2035 schedule [156]. This mass adoption is supported by the world's most extensive charging network, comprising 12.8 million charging piles and thousands of battery-swapping stations, with advanced fast-charging technologies already deployed at scale [152]

Despite this massive, state-driven scale-up in both the PV and EV sectors, China starkly exemplifies the SolarEV paradox, as its systemic Vehicle-to-Grid (V2G) integration remains in the nascent, pilot stages, fitting the 'Top-Down Strategic Planner' archetype. Although V2G is a national strategic priority, explicitly included in the New Energy Vehicle Industry Development Plan (2021-2035) and advanced via a "central policy guidance + local pilot projects" approach [114,157–159], progress is critically impeded by a "triple challenge". Technologically, the current national GB/T charging standard is unidirectional, creating a significant barrier pending revisions aimed at achieving compatibility with international standards like ISO15118-20 [160,161]. Economically, high equipment renovation costs and imperfect electricity pricing mechanisms have hindered the formation of viable business models. Socially, consumer acceptance remains low, primarily due to persistent concerns regarding battery degradation and a low willingness to participate in V2G programs.

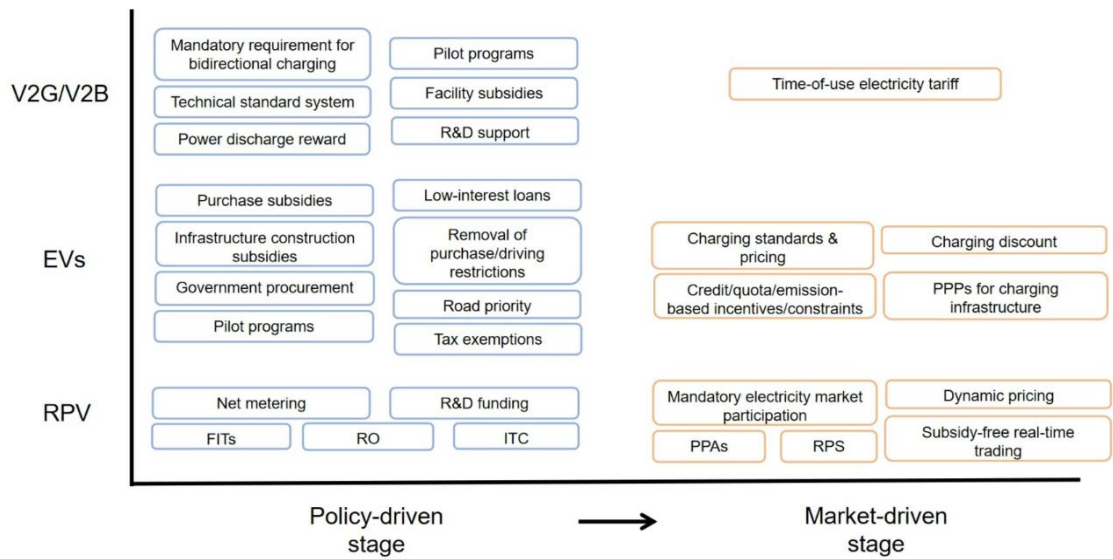

Figure 4. Summary of Chinese Policies. RO: Renewables Obligation PPAs: Power Purchase Agreements. RPS：Renewable Portfolio Standards. PPPs: Public-Private Partnerships. ITC: Investment Tax Credit. R&D: Research and Development. RPV: Rooftop PV.

5.3 France

France presents a unique national context for energy transition, characterized by a power system heavily reliant on low-carbon nuclear generation (providing ~67% of electricity in 2024), which shifts the primary decarbonization focus toward electrifying the transport and heating sectors [162]. The French EV market is relatively mature (16.9% market share in 2024) [163], supported by a $CO_2$-indexed incentive/penalty scheme, while rooftop PV deployment, although steadily growing (12.2 GW cumulative capacity in 2023) [164], remains a smaller contributor to overall electricity generation compared to nuclear.

Within this context, France distinguishes itself as a leading example of the 'Regulation-First Integrator' archetype, demonstrating successful early commercialization of V2G services despite potentially lower immediate grid balancing pressures compared to countries with higher variable renewable energy shares. The launch of one of Europe's first scalable commercial V2G service in 2024 (a partnership between Renault and aggregator The Mobility House) [165] was significantly enabled by strong policy coherence and favorable institutional arrangements: a supportive regulatory environment proactively addressing V2G, near-universal deployment of smart meters facilitating bidirectional data flow, and the presence of a single, dominant Distribution System Operator (Enedis) [166], which simplified technical coordination and standardization. This proactive approach, prioritizing the establishment of clear market rules and grid access protocols for aggregators, has significantly addressed major policy and economic barriers at an early stage, allowing market-driven innovation that is now beginning to emerge and is positioning France as a European frontrunner in V2G deployment [167].

5.4 India

India's energy transition is characterized by rapid, policy-driven growth in its core technology sectors, alongside ambitious national targets (e.g., 500 GW non-fossil capacity by 2030) [168]. The rooftop PV sector has expanded significantly, reaching 18.8 GW by June 2025, supported by major subsidy schemes like 'PM Surya Ghar' [169–171]. This growth, however, is predominantly concentrated among commercial and industrial users, with residential adoption only recently accelerating, and persistent challenges remain regarding policy consistency and an urban-rural access gap [169,172]. Concurrently, India's EV market is accelerating, exceeding 5.6 million registered vehicles by mid-2025 (8% new sales share in 2024), though this adoption is heavily dominated by two- and three-wheelers [173–175]. While supportive policies (e.g., FAME II) drive this trend, significant barriers include high upfront costs and dependence on imported batteries [175,176].

Despite this parallel growth in PV and EV adoption, India exemplifies the 'Nascent Adopter' archetype, where systemic integration through bidirectional charging remains almost entirely absent, highlighting the core SolarEV paradox. Implementation of V2G or V2H is confined to small-scale pilot studies, with no large-scale commercial deployments [177–179]. This stagnation is primarily attributable to foundational barriers in the policy and economic pillars. Critically, no comprehensive regulatory framework for commercial V2G operations exists; draft guidelines are pending, but there are no established compensation mechanisms, market access rules, ownership clarity, or taxation frameworks for grid services [180,181]. Currently, EVs and other DERs are not permitted to participate in wholesale or ancillary service markets, creating a fundamental economic barrier [182].

While significant policy, market, and social hurdles persist, nascent enablers are emerging. Policy alignment is gradually growing through initiatives like the "Solar City" mandates and domestic manufacturing incentives (Ministry of New and Renewable Energy, 2023; Regulatory Assistance Project, 2024a). The VPP ecosystem is evolving through small pilots [185,186], but aggregator remuneration and market access frameworks remain undefined [187]. A critical step forward is the national mandate for TOU tariffs by April 2025, which aims to create price signals for flexibility [188–190]. However, social barriers are substantial regarding battery degradation and warranties [191–194]. Ultimately, while India's vast solar potential [195] provides a strong foundation, scaling the SolarEV City concept necessitates the comprehensive development of its entire socio-technical framework.

5.5 Indonesia

Indonesia, a rapidly urbanizing archipelago, confronts the dual challenges of meeting high economic growth targets (e.g., 8% by 2029), while transitioning from a heavy fossil fuel dependency towards its Net Zero Emissions 2060 goal, as outlined in the new National Energy Policy (KEN) (Government Regulation No. 40 of 2025). Solar PV is identified as a main strategy to increase the New and Renewable Energy (NRE) share from 15.3% in 2024 to 70-72% by 2060 [196], supported by ambitious targets in the national utility's (PT PLN) plan (17.1 GW solar 2025-2034) and a new 100 GW presidential solar program. Despite this momentum, and growing rooftop PV adoption by industries seeking cost reductions and clean energy credentials, overall deployment remains extremely low, fitting the 'Nascent Adopter' archetype.

Bidirectional charging (V2G/V2H) is an emerging technology concept in Indonesia with significant theoretical potential to revolutionize grid operations by providing load balancing, peak shaving, and resilience as the EV fleet grows (Table 1). However, it is currently confined to academic modeling studies; field trials are needed [26]. This highlights the SolarEV paradox, as existing policies (e.g., Presidential Regulation No. 55/219; MEMR Regulation No. 13/2020) focus exclusively on incentivizing unidirectional EV charging infrastructure and do not address bidirectional capability, communication standards, or V2G interoperability. Tariff schemes similarly focus only on making charging attractive, with no mechanisms to compensate EV owners for discharging energy or providing grid services.

Realizing V2G's potential to support Indonesia's renewable expansion and defer grid investments requires addressing a comprehensive list of challenges across the socio-technical framework. These include: (i) technical readiness (lack of V2G-capable hardware, battery degradation concerns); (ii) a complete absence of policy and regulatory frameworks for grid feedback, metering, billing, and safety; (iii) infrastructure and grid constraints (inability to handle reverse flows); (iv) the lack of a viable economic business models (no compensation incentives); (v) low consumer acceptance (warranty and trust issues); and (vi) no interoperable standards. Given the policy momentum for EVs, V2G pilots may emerge in the next 2-5 years, but full-scale, commercially viable deployment remains a distant prospect contingent on comprehensive regulatory development.

5.6 Japan

Japan's national context is defined by ambitious decarbonization goals (2050 carbon neutrality, 100% electrified new car sales by 2035) [197] and significant land-use constraints, which elevate the strategic importance of rooftop PV [198]. Consequently, rooftop PV constitutes a major share of new capacity (68% in 2024) [6], and modeling confirms the high potential for integrated "SolarEV City" systems to achieve deep $CO_2$ emission reductions (54-95%) in this setting [7]. This context, however, frames a key national paradox: despite possessing a mature PV market, a pioneering auto industry, and proven technical feasibility for aggregation [199], domestic EV adoption remains exceptionally low (<3% market share in 2024) [200].

Japan distinctly embodies the 'Resilience-Driven Pioneer' archetype, where the trajectory of vehicle-centric energy integration has been uniquely shaped by non-market forces. The 2011 Fukushima disaster prioritized household-level disaster resilience, spurring coordinated government-industry action to standardize the bidirectional CHAdeMO protocol specifically for V2H applications (V2G capable as of 2025). This resulted in Japan establishing the world's most advanced commercial V2H market (approx. 28,000 units sold by end 2024), effectively leveraging parked EVs as backup power [201–203]. However, this success in V2H represents a specific form of fragmented integration; it has not translated into widespread V2G implementation, which remains largely confined to pilot projects with the broader V2G market still considered "generally limited" [167].

The supporting ecosystem for full integration is evolving. National strategies like Green Transformation (GX) and Digital Transformation (DX), alongside Smart City platforms and microgrid demonstrations [204], indirectly support SolarEV principles [87,205–207]. A sophisticated smart grid infrastructure [208] and emerging P2P trading pilots [209–211] provide a digital foundation. However, increasing grid congestion from high PV penetration underscores the urgent need for the flexibility V2G can provide [212]. The primary barrier remains the policy and market framework: regulatory discussions are now underway to define rules for EV participation in electricity markets, highlighting that the bottleneck for Japan's transition from V2H to V2G is no longer technological but regulatory [201].

5.7 Netherlands

The Netherlands demonstrates exceptionally high ambition in its energy transition, driven by stringent climate targets and spatial constraints favoring efficient resource utilization. It boasts Europe's highest installed PV capacity per capita (1,357 W/person), with a cumulative capacity of approximately 28.6 GW by 2024, resulting in solar energy accounting for nearly 18% of gross electricity production [213,214]. This rapid PV growth, spurred significantly by net metering policies, means roughly one-third of Dutch homes now host solar panels [215]. Concurrently, the EV market is a European leader, with over 1 million plug-in vehicles registered by 2025 (11.2% of the fleet) and new plug-in sales reaching 58% market share [216]. The anticipated abolition of net metering creates strong economic pressure for increasing self-consumption of solar energy, positioning EV charging and V2X integration as viable solutions alongside stationary storage, although managing drastic seasonal variation remains a key challenge [217].

Fitting the 'Regulation-First Integrator' archetype with a strong emphasis on technological innovation, the Netherlands serves as a crucial European "living lab" for V2X and solar integration, pursuing two key development pathways. Firstly, TU Delft, in collaboration with industry partners, developed and demonstrated a 10kW integrated direct DC solar EV charger (Fig. 5a). This system enables direct DC-to-DC power transfer from PV panels to the EV and incorporates a bidirectional offboard V2G charger, minimizing conversion losses and potentially reducing system costs [218,219]. This architecture can also be expanded to include local stationary battery storage (Fig. 5b). Secondly, significant progress is being made using onboard bidirectional AC chargers within EVs, compatible with the ISO 15118-20 standard. The city of Utrecht, notably, is deploying V2G services using specialized Renault and Hyundai vehicles equipped with such chargers, representing a potentially more cost-effective and scalable model requiring only a compatible AC charging station [220,221]. A key challenge for this onboard AC approach, however, remains ensuring compliance with diverse national grid codes, currently managed through geofencing and ongoing regulatory harmonization efforts.

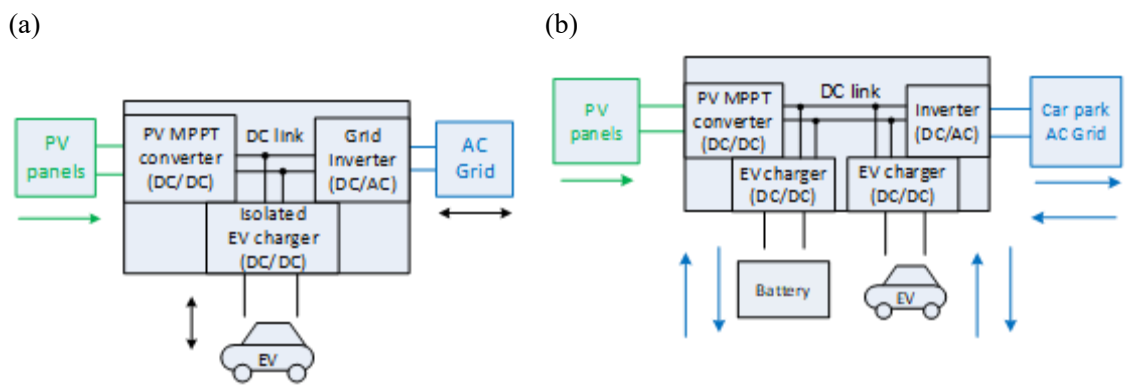

Figire 5. (a) 10kW integrated direct DC offboard solar EV charger that functions both as solar inverter and V2G charger; (b) with possibility to add a local battery to provide energy storage when EV is not available.

5.8 South Korea

South Korea's national strategy for deep decarbonization, formalized by its 2050 carbon neutrality commitment [222], relies on aggressive, parallel targets to transform the urban energy landscape. This includes ambitious goals for both EV adoption (aiming for 1.13 million EVs by 2025) [223] and solar PV expansion (targeting 55.7 GW by 2030) [224]. This top-down national push is mirrored by highly ambitious city-level initiatives. Seoul's "Solar City Seoul" program, for instance, targeted 1GW of rooftop PV capacity [225,226], complemented by a massive EV charging rollout aiming for 220,000 chargers by 2026 to ensure 5-minute access for all residents, sometimes integrating solar generation directly at charging hubs [227–229]. Other national smart-city pilots, such as Busan's Eco Delta City and Jeju's Carbon-Free Island 2030, further showcase this commitment to integrated, technology-forward urban development [230,231]. Modeling studies confirm the high potential of this approach, projecting that PV-EV integration could cut Seoul's urban $CO_2$ emissions by approximately 49% by 2030 [232].

This approach fits a hybrid model, combining 'Top-Down Strategic Planning' with strong 'corporate-led technological innovation'. Systemic integration is actively pursued through policies supporting synchronized deployment, smart grids, and incentive structures like TOU tariffs and demand response programs [232]. V2G and V2B technologies, supported by AI/IoT platforms, are in advanced pilot phases, often leveraged by major domestic corporations within these smart city frameworks [230,233]. Despite strong public support for clean energy [233], translating these pilots into a scaled, commercially viable national market exposes the SolarEV paradox. The primary social barrier is structural: ensuring equitable charging and PV access for the large proportion of the population residing in high-rise apartments. While this is being addressed through targeted subsidies and innovative solutions like lamppost chargers [229], it highlights the significant friction between ambitious top-down goals and the socio-spatial realities of urban deployment. Concurrently, new building mandates, such as Zero Energy Building (ZEB) standards effective June 2025, introduce additional implementation costs [234].

5.9 United States

The United States energy transition unfolds within a complex and highly fragmented governance landscape, shaped by an interplay of federal policies, diverse state-level mandates, and significant market forces. Rooftop PV capacity has grown rapidly from 7.3GW in 2014 to 39.5 GW in 2022, constituting roughly one-third of total national solar capacity [235]. This growth is substantially driven by federal incentives, such as the Inflation Reduction Act's 30% tax credit [236] (though its longevity is subject to political change [237]), and highly variable state-level incentives, alongside net metering and clear consumer cost savings [238–240]. Concurrently, the EV market is expanding, achieving 10% market share for new sales in 2024 [241] and supported by federal infrastructure programs like NEVI [242]. However, overall fleet penetration remains low (~1.4%) [243], and adoption faces persistent barriers including cost, charging times, and infrastructure availability [244,245].

This context positions the US predominantly within the 'Market-Led, Fragmented Innovator' archetype, where the SolarEV paradox is particularly pronounced. Technological innovation from the private sector is vibrant; automakers (e.g., Ford F-150 Lightning, Kia EV9) are commercially launching vehicles with V2H capabilities, primarily marketed as residential backup power solutions. V2H is comparatively mature as it is a more straightforward, behind-the-meter application requiring specialized home installation. In stark contrast, systemic V2G integration, which necessitates complex coordination between automakers, utilities, and regulators, remains nascent and largely confined to small-scale pilot programs, such as Baltimore Gas and Electric's residential V2G trial [246,247].

The primary barrier to scaling from isolated V2H products to systemic V2G integration is the significant fragmentation across the policy and economic pillars. The US lacks a unified national regulatory framework for V2G interconnection and market participation, requiring navigation of over 3,000 distinct utilities, each with different rules and operational models. This governance fragmentation, combined with inconsistent state-level policies and historical delays in the full implementation of bidirectional capabilities within the dominant Combined Charging System (CCS) protocol (ISO 15118; SAE J3072), creates significant structural barriers to achieving scale and interoperability. Consequently, progress remains piecemeal and geographically concentrated, awaiting greater federal policy coherence and market structure reform to bridge the integration gap.

5.10 Cross-Case Synthesis: Divergent Development Archetypes

The comparative analysis conducted across these nine countries confirms that nations are not converging towards a single, uniform model for SolarEV City development. Instead, distinct development archetypes emerge, shaped by specific national context, policy priorities, and existing institutional structures (Fig. 6):

1. The Resilience-Driven Pioneer (Japan): This archetype is characterized by a primary policy focus on disaster preparedness and energy security at the household level. This specific driver spurred early, coordinated government-industry action to standardize and commercialize V2H technology [248], resulting in the world's most mature V2H market. However, this success in household resilience has not translated proportionally into systemic grid integration (V2G), which lags due to different regulatory and economic barrier [7].
2. The Market-Led, Fragmented Innovator (U.S., Australia): This pathway is characterized by dynamic, bottom-up innovation, often from automakers and tech startups, and strong consumer-led demand (e.g., high rooftop PV uptake in Australia) [136]. However, progress toward systemic integration is chaotic and stymied by significant fragmentation in the policy and regulatory landscape, a lack of national coordination, a complex multi-utility environment (especially in the U.S.), and delays in the adoption of key technical standards [145].

3. The Top-Down Strategic Planner (China, South Korea): In this model, V2G integration is explicitly incorporated into national industrial and energy strategies, driven by strong central government direction [114] or championed by major domestic corporations aligned with national goals [222]. Development often proceeds through large-scale, state-sanctioned pilot projects (e.g., Busan Eco Delta City), but can be constrained by the pace of centralized standard-setting (e.g., GB/T standard in China) or market design finalization.

4. The Regulation-First Integrator (France, Netherlands): This archetype is defined by proactive, coherent, and often national-level regulatory frameworks that create a stable and predictable environment for investment. Factors such as smart meter rollouts and streamlined grid operator structures (e.g., a single dominant Distribution System Operator (DSO) in France), and a culture of "living lab" experimentation (e.g., Utrecht and TU Delft in the Netherlands) [218] enable a smoother transition from pilot projects to early commercial V2G service deployment.

5. The Nascent Adopter (India, Indonesia): These countries are characterized by immense future potential but significant foundational barriers across the socio-technical system. In these countries, the primary policy focus remains on individually accelerating the basic adoption of rooftop PV and EVs. Systemic integration through V2G is acknowledged as a long-term objective but is contingent on major grid modernization and policy development.

These distinct archetypes demonstrate that there is no one-size-fits-all pathway to a SolarEV City. The specific challenges and opportunities are deeply embedded in local institutional and technical contexts, underscoring the critical need for tailored, context-specific strategies rather than the wholesale importation of models from one region to another.

| Characteristic | Policy Focus | Key Enablers | Key Barriers | V2G Integration |
|---|---|---|---|---|
| Resilience-Driven Pioneer (Japan) | Disaster preparedness, energy security | Standardized V2H technology, early government-industry action | Systemic grid integration lags | Lags due to regulatory and economic barriers |
| Market-Led, Fragmented Innovator (U.S., Australia) | Consumer demand, bottom-up innovation | Dynamic innovation, strong consumer demand | Policy and regulatory fragmentation, lack of national coordination | Chaotic and stymied by fragmentation |
| Top-Down Strategic Planner (China, South Korea) | National industrial and energy strategies | State-sanctioned pilot projects, strong central government direction | Pace of centralized standard-setting, market design finalization | Explicitly incorporated into national strategies |
| Regulation-First Integrator (France, Netherlands) | Proactive, coherent regulatory frameworks | Smart meter rollouts, streamlined grid operator structures | Transition from pilot projects to commercial deployment | Smoother transition from pilot projects |
| Nascent Adopter (India, Indonesia) | Basic adoption of PV and EVs | Grid modernization, policy development | Foundational barriers across the socio-technical system | Long-term objective, contingent on grid modernization |

Figure 6. SolarEV City development archetypes.

## 6. Challenges and Future Perspectives

Despite the compelling theoretical potential and advancing technological capabilities, transitioning the SolarEV City concept from niche pilots to widespread urban reality confronts significant, intertwined challenges across technical infrastructure limits, economic viability hurdles, policy and governance inertia, and critical social equity dimensions. Addressing these challenges requires a concerted multi-faceted approach.

6.1 Technical and Infrastructural Barriers

6.1.1 Distributed-Grid Hosting Capacity

A primary technical impediment to scaling SolarEV City concept is the limited hosting capacity (HC) of existing urban distribution grids, defined as the maximum DER injection a network can accommodate without violating thermal, voltage, or power quality constraints [249]. Legacy networks, designed for unidirectional power flow, face compounded stresses from this framework, which integrates high-penetration PV and EVs across diverse residential, commercial, and industrial zones [15,250]. These stresses include reverse power flows from midday PV surpluses [251,252] and dynamic, spatiotemporal load shifts created by EV mobility patterns (e.g., consumer charging), which differ significantly from stationary storage [253]. Studies project that unmanaged EV charging alone could necessitate costly, large-scale capacity upgrades on most feeders [253].

Research into mitigation strategies highlights the critical role of intelligent coordination. While basic smart charging offers moderate benefits [249], studies examining concurrent high PV and EV penetration confirm that uncontrolled charging significantly exacerbates evening peak demand and network stress [254,255]. Conversely, coordinated strategies, particularly those leveraging V2G capability and advanced inverter functions, are shown to effectively mitigate thermal and voltage challenges, potentially restoring HC to levels comparable with PV-only scenarios [256–260]. Despite this progress, critical research gaps remain; existing models often analyze PV and EV impacts in isolation, rely on synthetic network topologies, and fail to integrate DER scheduling under uncertainty. Future research must move beyond static HC analysis and develop dynamic models that treat HC as a variable co-optimized with V2G/VPP economic dispatch, quantifying how market and policy mechanisms (e.g., P2P trading) can expand effective grid capacity.

6.1.2 Battery Lifecycle, Recycling, and Material Constrains

The viability and sustainability of the SolarEV City concept are intricately dependent on EV battery performance, longevity, and end-of-life management. Utilizing the vast, aggregated storage capacity of the EV fleet offers significant systemic flexibility, potentially reducing the need for massive investments in new, dedicated stationary batteries to balance grid intermittency. However, this strategy presents two major systemic challenges. First, the massive scale-up of EVs required exerts significant pressure on global supply chains for critical minerals (lithium, cobalt, etc.), necessitating robust, efficient, and economically viable pathways for battery recycling and second-life applications as a critical prerequisite for long-term sustainability. Second, EV batteries (e.g., 20-100 kWh) are insufficient for managing deep, long-duration seasonal variations in solar generation, an issue particularly acute in high-latitude regions; for example, PV generation in the Netherlands can vary by up to 75 times between a sunny day in summer to a cloudy day in winter, with an average variation of about 5 times between the months [217].

At the vehicle level, a primary barrier to societal acceptance remains consumer concern over accelerated battery degradation, amplified by the fact that most manufacturer warranties do not explicitly cover V2G use [43,124] with exceptions in Japan where some automakers allow the use of V2H within their warranty [261]. This relationship is complex: while V2X increases cyclic aging through repeated use, intelligent V2G algorithms can simultaneously reduce calendar aging by actively managing the battery's state-of-charge (SoC) to avoid high-SoC/high-temperature conditions that accelerate degradation [262]. Consequently, the net effect of this trade-off may be neutral, or in some optimized scenarios, even reduce overall degradation [125−128]. This outcome is highly dependent on battery chemistry (with LFP generally considered more suitable than NMC), the V2X service parameters (C-rate, DoD, frequency), and effective thermal management [262]. Future work must bridge the gap between engineering-centric battery degradation models [125–128] and the socio-economic reality of consumer acceptance. A key agenda is developing and validating 'V2G-ready' warranties, translating technical degradation mitigation strategies into bankable and trustworthy assurances for users.

6.1.3 Implications on urban design

The SolarEV City concept demands a transformative approach to urban design, moving beyond traditional planning paradigms to fundamentally integrate energy infrastructure into the urban fabric. This necessitates a total reconsideration of 3D urban structures [263], including the distribution of buildings and their usages, to optimize PV energy generation and consumption and thereby maximize self-sufficiency within the city [264]. Urban layouts must be reconfigured to prioritize this infrastructure while maintaining livability, such as through dual-use solar canopies [265]. Similarly, the ubiquitous integration of EV charging infrastructure must be carefully designed to ensure accessibility and avoid spatial conflict in dense pedestrian areas [130]. Furthermore, this integration requires a shift from static planning to dynamic, future-proof design principles; urban environments must become modular and scalable to adapt to the rapid evolution of technologies like flexible BIPV and V2G systems, ensuring long-term resilience [265–267]. Advanced modeling tools, such as 3D urban simulations, are essential for evaluating the complex impacts of these interventions on traffic, land use, and urban functionality [267].

This paradigm shift in urban design must also explicitly embed environmental and social co-benefits through a systems-thinking approach, integrating PV/EV infrastructure with green spaces to enhance air quality and mitigate urban heat [268]. A critical design challenge within this framework is ensuring energy justice and equitable access to SolarEV benefits. As marginalized communities often lack access to sustainable infrastructure [130], urban design must prioritize deployment in underserved areas, utilizing participatory processes to incorporate resident input and foster social cohesion [131]. Successfully navigating these complex implications requires deep interdisciplinary collaboration (integrating energy engineering, social policy, and planning) and may be aided by advanced simulation tools like digital twins, enabling data-driven decisions to create urban environments that are not only energy-efficient but also inclusive and resilient [269].

6.2 Economic, Market, and Social Hurdles

While technical barriers are being addressed, significant socio-economic challenges remain paramount. Developing viable and scalable business models for V2G services that deliver clear value propositions to all stakeholders, EV owners (compensating for battery degradation and inconvenience), aggregators (ensuring profitability), and grid operators (providing reliable, cost-effective services) has been proven to being complex. Incumbent utility business models, often established on centralized generation and volumetric energy sales, may face disruption from highly decentralized VPPs and P2P trading, potentially leading to institutional resistance or slow adaptation of market rules. Critically, the inherent "equity gap" associated with the high upfront costs of participation risks creating a socially stratified energy system, demanding proactive policy interventions focused on inclusive financing, community-based models, and equitable benefit distribution to ensure a just transition [42,131]. Building broad social acceptance necessitates effectively addressing consumer behavioral barriers through trusted information, user control, and demonstrated reliability.

6.3. Governance, Policy, and International Collaboration

Regulatory fragmentation and policy inertia constitute the primary bottlenecks impeding scaled deployment [270]. While national-level policies, such as Japan's 2050 carbon neutrality goal or the US Inflation Reduction Act, create crucial landscape-level pressure on existing socio-technical regimes, national governments often lack the context-specific knowledge and multi-stakeholder connections to steer complex urban transitions alone [271]. Realizing the SolarEV City vision therefore requires a paradigm shift in urban governance towards integrated planning [272], overcoming sectoral fragmentation across energy, transportation, and land-use authorities. This includes updating building codes for solar PV and EV readiness, reforming zoning for distributed infrastructure, and fundamentally reorienting utility planning processes to embrace DERs as core grid resources.

Overcoming these barriers requires more than just top-down mandates; it necessitates "strategic niche management" to cultivate the "PV + EV" concept [273]. This is often achieved through "bottom-up" multi-stakeholder collaboration, as demonstrated in case studies like the Kyoto Miraimon Project [270]. Such platforms function to align the diverse, and often conflicting, interests of key actors, including industry (e.g., automakers, utilities), academia, NGOs, and local government, around a shared, scientifically grounded vision. Universities, in particular, can play a crucial role as "niche intermediaries", forming platforms, providing impartial techno-economic analysis to demonstrate the potential, and facilitating social experiments to identify and overcome context-specific barriers [270,274]. These action-oriented, transdisciplinary processes are vital for building trust, co-creating solutions, and providing iterative feedback to decision makers, thereby facilitating the co-evolution of technology and institutions required for a sustainability transition.

Finally, as the challenges of scaling the SolarEV City concept are global in nature, enhanced international collaboration is imperative. This includes concerted efforts by standards organizations to harmonize bidirectional charging communication protocols (e.g., ISO 15118-20 and ChoJi) to foster a seamless, interoperable global market. Furthermore, establishing formal mechanisms for governments, regulators, and platform facilitators to actively share best practices and policy learnings from diverse national V2G experiments is essential to accelerate the collective learning curve. A coordinated international research agenda is also needed to address common, high priority challenges, such as developing open source models for hosting capacity analysis, conducting longitudinal studies on battery degradation, and establishing common methodologies for assessing the social equity impacts of this transition.

## 7. Conclusion

The *SolarEV City Concept*, built on the synergistic integration of ubiquitous rooftop PV generation with EV fleets as decentralized storage, offers a technically feasible and powerful pathway for deep urban decarbonization. This review has critically examined the central paradox defining the current state of progress: the rapid, uncoordinated market growth of PV and EVs starkly contrasts with the slow, fragmented realization of their systemic integration. Applying a socio-technical framework, our analysis concludes that the primary impediments are no longer rooted in technological feasibility but in critical misalignments across other domains: (1) the lack of globally interoperable bidirectional charging standards (the "protocol challenge"); (2) the absence of viable V2G business models, often facing incumbent resistance; (3) persistent regulatory gaps hindering market access and failing to address grid constraints; and (4) profound, unaddressed social equity challenges that risk a non-inclusive transition. The emergence of divergent national archetypes, from Japan's resilience-driven V2H market to Europe's regulation-first commercial V2G, confirms that these socio-technical barriers now shape development. Ultimately, realizing the transformative potential of the SolarEV City concept necessitates a fundamental paradigm shift: moving beyond the promotion of discrete technologies toward the strategic, multi-stakeholder orchestration of an integrated, equitable, and resilient urban energy system. In conclusion, the SolarEV City Concept represents more than a technical solution—it is a new paradigm for urban energy governance rooted in decentralization, flexibility, and citizen empowerment. If adopted with the necessary foresight and coordination, it could become a cornerstone of global efforts to build just, resilient, and deeply decarbonized urban futures.

**CRediT authorship contribution statement**
T. Kobashi: Writing – review & editing, Writing – original draft, Visualization, Validation, Resources, Project administration, Methodology, Investigation, Funding acquisition, Formal analysis, Conceptualization. R. C. Mouli: Writing – review & editing, Writing – original draft, Visualization, Validation. J. Liu: Writing – review & editing, Writing – original draft, Visualization, Validation. S.

Chang: Writing – review & editing, Writing – original draft, Validation. C. D. Harper: Writing – review & editing, Writing – original draft, Validation. R. Zhou: Writing – review & editing, Writing – original draft, Validation. G. R. Dewi: Writing – review & editing, Writing – original draft, Validation. U. W. R. Siagian: Writing – review & editing, Writing – original draft, Validation. J. Kang: Writing – review & editing, Writing – original draft, Validation. G. P. Patankaar: Writing – review & editing, Writing – original draft, Validation. Z. H. Rather: Writing – review & editing, Writing – original draft, Validation. K. Say: Writing – review & editing, Writing – original draft, Validation. K. Tanaka: Writing – review & editing, Validation. P. Cias: Writing – review & editing, Validation. D. M. Kammen: Writing – review & editing, Validation.

**Declaration of competing interest**

The authors declare that they have no known competing financial interests or personal relationships that could have appeared to influence the work reported in this paper.

**Acknowledgements**

This work was supported by the KAKENHI (23K11520) and FREI.

**Data availability**

Data will be made available on request.

[222] The Government of the Republic of Korea. 2050 Carbon Neutral Strategy of the Republic of Korea: Towards a sustainable and green society. 2020.

[223] Electrek. South Korea to boost EV numbers to 1.13 million by 2025 2020. https://electrek.co/2020/09/08/south-korea-to-boost-ev-numbers-to-1-13-million-by-2025/ (accessed June 30, 2025).

[224] The Korea Times. Economic downturn, aging population push up youth unemployment rate 2018.

[225] Seoul Metropolitan Government. Seoul to Introduce Diverse EV Chargers 2022. http://english.seoul.go.kr/seoul-to-introduce-diverse-ev-chargers/ (accessed June 30, 2025).

[226] Seoul Metropolitan Government. One in three Houses in Seoul to Have Photovoltaic Facility 2017. http://english.seoul.go.kr/one-three-houses-seoul-photovoltaic-facility/ (accessed June 30, 2025).

[227] Fast Company. Seoul plans to install more than 200,000 EV chargers by 2026 2022. https://www.fastcompany.com/90790989/seoul-plans-to-install-more-than-200000-ev-chargers-by-2026 (accessed June 30, 2025).

[228] Kim E, Heo E. Key drivers behind the adoption of electric vehicle in Korea: An analysis of the revealed preferences. Sustainability (Switzerland) 2019;11. https://doi.org/10.3390/su11236854.

[229] The Korea Herald. Vehicle-to-grid service approved under regulatory sandbox 2021. https://www.koreaherald.com/article/2661894 (accessed June 30, 2025).

[230] Han D, Kim JH. Multiple Smart Cities: The Case of the Eco Delta City in South Korea. Sustainability (Switzerland) 2022;14. https://doi.org/10.3390/su14106243.

[231] Invest Korea. Jeju Aims to be Carbon Free by 2030 2018. https://www.investkorea.org/ik-en/bbs/i-5045/detail.do?ntt_sn=478142 (accessed June 30, 2025).

[232] Han G, An YS, Kim JK, Jung DE, Joo HJ, Kim H, et al. Analysis of grid flexibility in 100% electrified urban energy community: A year-long empirical study. Sustain Cities Soc 2024;113. https://doi.org/10.1016/j.scs.2024.105648.

[233] Kim JH, Kim SY, Yoo SH. Public acceptance of the "renewable energy 3020 plan": Evidence from a contingent valuation study in South Korea. Sustainability (Switzerland) 2020;12. https://doi.org/10.3390/SU12083151.

[234] Maeil Business Newspaper. Government enforces mandatory solar energy standards, pushing up housing costs 2025. https://m.mk.co.kr/news/realestate/11346732 (accessed June 30, 2025).

[235] EIA. Record U.S. small-scale solar capacity was added in 2022 2023. https://www.eia.gov/todayinenergy/detail.php?id=60341#:~:text=Many%20of%20t